# Switching of intra-orbital spin excitations in electron-doped iron pnictide superconductors

Soshi Iimura[1], Satoru Matsuishi[2], Masashi Miyakawa[3], Takashi Taniguchi[4], Katsuhiro Suzuki[5], Hidetomo Usui[5], Kazuhiko Kuroki[6], Ryoichi Kajimoto[7], Mitsutaka Nakamura[8], Yasuhiro Inamura[8], Kazuhiko Ikeuchi[7], Sungdae Ji[7] and Hideo Hosono[1,9*]

[1]*Materials and Structures Laboratory, Tokyo Institute of Technology, 4259 Nagatsuta-cho, Midori-ku, Yokohama 226-8503, Japan*
[2]*Materials Research Center for Element Strategy, Tokyo Institute of Technology, 4259 Nagatsuta-cho, Midori-ku, Yokohama 226-8503, Japan*
[3]*Superconducting Properties Unit, National Institute for Materials Science, 1-1 Namiki, Tsukuba, Ibaraki 305-0044, Japan*
[4]*Materials Processing Unit, National Institute for Materials Science, 1-1 Namiki, Tsukuba, 305-0044, Japan*
[5]*Department of Engineering Science, The University of Electro-Communications, Chofu, Tokyo 182-8585, Japan*
[6]*Department of Physics, Osaka University, 1-1 Machikaneyama Toyonaka, Osaka 560-0043, Japan*
[7]*Research Center for Neutron Science and Technology, Comprehensive Research Organization for Science and Society (CROSS), Tokai, Ibaraki 319-1106, Japan*
[8]*J-PARC Center, Japan Atomic Energy Agency, Tokai, Ibaraki 319-1195, Japan*
[9]*Frontier Research Center, Tokyo Institute of Technology, 4259 Nagatsuta-cho, Midori-ku, Yokohama 226-8503, Japan*

We investigate the doping dependence of the magnetic excitations in two-superconducting-dome-system LaFeAsO$_{1-x}$D$_x$. Using inelastic neutron scattering, spin fluctuations at different wavenumbers were observed under both superconducting domes around $x$ = 0.1 and 0.4, but vanished at $x$ = 0.2 corresponding to the $T_c$ valley. Theoretical calculations indicate that the characteristic doping dependence of spin fluctuations is rationally explained as a consequence of the switching of the two intra-orbital nestings within Fe-3$d_{YZ, ZX}$ and 3$d_{X2-Y2}$ by electron doping. The present results imply that the multi-orbital nature plays an important role in the doping and / or material dependence of the $T_c$ of the iron pnictide superconductors.

Since the discovery of superconductivity in LaFeAsO$_{1-x}$F$_x$ with $T_c$ = 26 K [1], iron pnictides have attracted great interest as a non-cuprate high-$T_c$ superconductor. By substituting other lanthanide ions (*Ln*) at La sites, maximum $T_c$ increased to 55 K [2]. The parent compound LaFeAsO, so-called 1111-type, undergoes a structural transition around 150 K followed by a paramagnetic-antiferromagnetic (AFM) transition [3,4]. If these transitions are suppressed by carrier doping or applied pressure, superconductivity emerges. As for the superconducting mechanism, a fully-gapped sign-reversing *s* wave state (*s*+− wave state) has been proposed based on nesting-related, spin density wave (SDW)-type fluctuations near a wavevector connecting the cylindrical hole and electron pockets [5,6]. This mechanism helped successfully to explain the suppression of superconductivity, namely the superconducting dome width in various compounds, such as LaFeAsO$_{1-x}$F$_x$, Ba(Fe$_{1-x}$Co$_x$)$_2$As$_2$ and NaFe$_{1-x}$Co$_x$As, upon electron doping up to the filling level of hole pockets ($e^−$/Fe ~ 0.1-0.2) [1,7,8].

However, very recently, Iimura *et al*. discovered a two-dome structure in the superconducting phase diagram of LaFeAsO$_{1-x}$H$_x$ in which to a large extent the hydride, instead of fluoride, substitutes oxygen as an electron dopant (Fig. 1) [9]. The first superconducting dome with the optimal $T_c$ of 29 K appears after the AFM phase is suppressed in almost the same manner as the F⁻-doped samples. The second dome with an optimal $T_c$ of 36 K is broadly spread around $e^−$/Fe ~ 0.35; a *T*-linear resistivity, often referred to as the non-Fermi liquid state, is observed above the $T_c$. Furthermore, substitution of other *Ln* ions at the La site assists the two domes to merge into a single, large dome with optimal $T_c$ of 55 K [10,11]. These findings suggest that investigating spin excitations under the newly found second dome, as well as the first dome in the LaFeAsO$_{1-x}$H$_x$, is crucial in understanding the superconductivity in the 1111-type iron pnictides. According to previous theoretical calculations, the

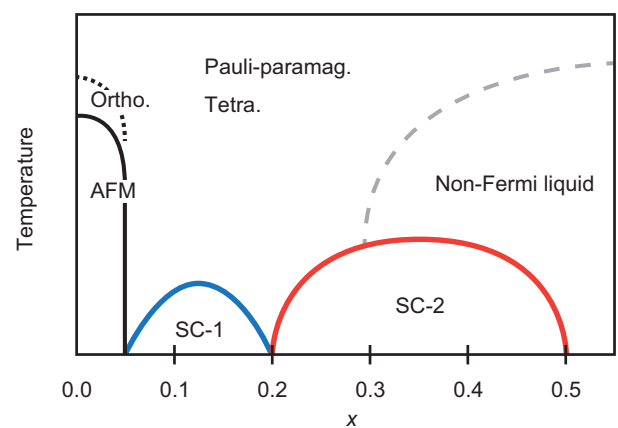

FIG. 1 (color) Schematic phase diagram of LaFeAsO$_{1-x}$H$_x$. Dotted and solid black lines show structural and AFM transitions, respectively; blue and red solid lines indicate $T_c$. The non-Fermi liquid states defined from the *T*-linear dependence of resistivity spread out for $x$ > 0.3 [9].

development of the effective spin fluctuations for superconductivity is unlikely in such heavy electron-doping conditions, $e^-$/Fe > 0.3 [12], whereas the observed $T$-linear resistivity indicates the crucial role of this mechanism [13].

Thus, we performed inelastic neutron scattering measurements, which can probe directly the imaginary part of the dynamical spin susceptibility [$\chi''(Q, E)$], to investigate the dynamic spin states under the superconducting dome in LaFeAsO$_{1-x}$H$_x$. Chosen were the chemical compositions $x$ = 0.1, 0.2, and 0.4 corresponding to the top of the first dome, the $T_c$ valley, and the top of the second dome, respectively. About 30 g of polycrystalline LaFeAsO$_{1-x}$D$_x$ were prepared for each $x$ by solid state reaction at 1323–1373 K, under pressure of 2 GPa by using a belt-type high pressure apparatus with a bore diameter of 60 mm. Deuterium (D) is used in samples to avoid large incoherent scattering from hydrogen. However, due to the insufficient deuterization, hydrogen (H) at oxygen site of LaFeAsO$_{1-x}$D$_x$ was detected as high-frequency phonons at energy-transfer $E \sim 130$ meV. It would come from the external deuterium source NaBD$_4$ with 10atom% of H in the high pressure cell. For the sample compositions $x$ = 0.1, 0.2, and 0.4, the $T_c$'s were 27, 14, and 34 K respectively, and the shielding volume fractions, determined by a SQUID magnetometer under a magnetic field of 10 Oe, were 20, 8, and 41% at $T$ = 2 K respectively. Several inelastic neutron scattering measurements were performed using the Fermi chopper spectrometer 4SEASONS in J-PARC using multi-incident energies of 152.5, 45.1, 21.5, and 12.5 meV at 7 K. All data presented here were obtained at an incident energy of 45.1 meV. For the $x$ = 0.4 sample, a high-temperature measurement at 150 K was also performed. To remove background noise, inelastic neutron scattering from an empty aluminium sample holder was measured at 7 K. Measuring times for each sample and the sample holder were 27 hours and 18 hours respectively, at a beam power of 210 kW.

Figure 2(a)-(c) show the doping dependency of the neutron inelastic scattering intensity from LaFeAsO$_{1-x}$D$_x$ with $x$ = 0.1, 0.2 and 0.4 at 7 K as a function of momentum transfer ($Q$) at several fixed energy transfers ($E$). For composition $x$ = 0.1 [Fig. 2(a)], a peak is observed at $Q$ = 1.1 Å$^{-1}$ and in 10 < $E$ < 17 meV. Gaussian fitting with a sloped background gives a peak position of $Q$ = 1.14 Å$^{-1}$, which is very close to the wavevector for the 2-dimensional stripe-type AFM order in the parent phase, $Q^{2D}_{AFM} = (1\ 0\ 0) \sim 1.1$ Å$^{-1}$ in orthorhombic notation, indicating fluctuations of the SDW [5]. For the $x$ = 0.2, the magnetic peak completely disappears [Fig. 2(b)]. Because the $T_c$ of 14 K is lower than that for the sample $x$ = 0.1, one would expect that assuming an enhanced spin excitation at $T_c$ the magnetic peak would move to lower energy; however, no peak was observed. As $x$ increased to 0.4 [Fig. 2(c)], a peak appears again at $Q$ = 1.25-1.38 Å$^{-1}$ in 16 < $E$ < 22 meV. To distinguish the magnetic peak from the nuclear peak, the momentum scan at $T$ = 150 K is shown for comparison [Fig. 2(d)]. Because the phonon scattering cross-section increases with temperature as $n(\omega) + 1$, where $n(\omega)$ is the Bose factor $(\exp^{\hbar\omega/kT} - 1)^{-1}$, the peak intensity should increase if its source is phonons. However, the peak observed at $T$ = 7 K disappears at $T$ = 150 K, indicating indeed magnetic excitations as source.

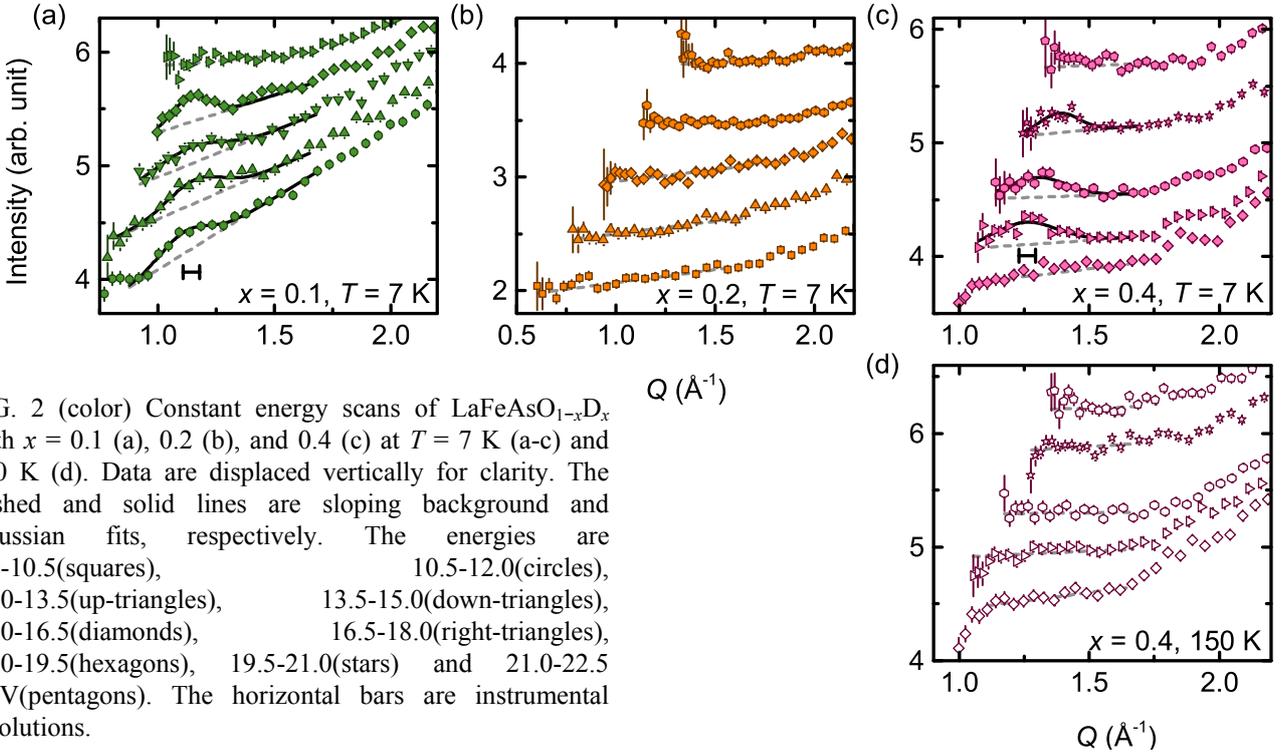

FIG. 2 (color) Constant energy scans of LaFeAsO$_{1-x}$D$_x$ with $x$ = 0.1 (a), 0.2 (b), and 0.4 (c) at $T$ = 7 K (a-c) and 150 K (d). Data are displaced vertically for clarity. The dashed and solid lines are sloping background and Gaussian fits, respectively. The energies are 9.0-10.5(squares), 10.5-12.0(circles), 12.0-13.5(up-triangles), 13.5-15.0(down-triangles), 15.0-16.5(diamonds), 16.5-18.0(right-triangles), 18.0-19.5(hexagons), 19.5-21.0(stars) and 21.0-22.5 meV(pentagons). The horizontal bars are instrumental resolutions.

Figure 3(a) shows the energy dependence of dynamical spin susceptibility $\chi_s''(E)$ calculated by integrating the fitted results of Fig. 2(a) and (c) over $Q$. The spin excitation spectra at $x = 0.1$ and 0.4 under superconducting states have peaks at $E = 13$ meV and 17 meV, respectively. The maximum $\chi_s''(E)$ value at $x = 0.4$ is larger than that at $x = 0.1$, which mainly comes from the difference of the magnetic moment on Fe, $p_{Fe} = 0.32$ and 0.85 $\mu_B$, respectively [14,15]. For unconventional superconductors covering various material systems, a simple quantitative relation $E_R/k_BT_c \sim 4$–6 ($E_R$: resonance energy) was valid [16] [Fig. 3(b)]. If we use $E_R = 13$ and 17 meV for the samples with $x = 0.1$ and 0.4 ($T_c = 27$ and 34 K), then values $E_R/k_BT_c \sim 5.6$ and $\sim 5.8$ agree well with this relation. The previous theoretical calculations have revealed that the form of $\chi_s''(E)$ is very sensitive for the pairing symmetry of the superconducting state [17-19]. The $d$ wave state produces a very weak resonant peak, whereas $s+-$ and $s++$ wave states lead to peaked $\chi_s''(E)$ at the energy of superconducting gap. The peaked structures of $\chi_s''(E)$ at $E \sim 6\ k_BT_c$ may suggest that the pairing symmetries of the two superconducting phases are the $s+-$ or $s++$ wave states.

In electron-doped iron pnictides, the development of spin fluctuations with $Q^{2D}_{AFM} \sim 1.1$ Å$^{-1}$ at $e^-$/Fe $\sim 0.1$ and its suppression at $e^-$/Fe $\sim 0.2$ were also observed in Ba(Fe$_{1-x}$Co$_x$)$_2$As$_2$ and LaFeAsO$_{1-x}$F$_x$ [20,21]. The concurrence is successfully explained by the doping dependence of the Fermi-surface nesting between hole and electron pockets, *i.e.*, electron doping monotonically suppresses spin fluctuations through the deterioration in nesting, which finally disappears around $e^-$/Fe $\sim 0.2$. The continual suppression of nesting, however, cannot account for the revival of spin fluctuations above $e^-$/Fe $> 0.3$. To investigate why and how these spin fluctuations develop and whether it induces superconductivity, we performed theoretical calculations based on the random phase approximation (RPA) applied to a five-orbital model of LaFeAsO$_{1-x}$H$_x$ that was derived from the first principles band calculation [22] exploiting the maximally localized Wannier orbitals [23,24]. The electron doping was modelled by the virtual crystal approximation (VCA) that assumes oxygen ($Z = 8$) behaves as a virtual atom with fractional nuclear charge ($Z = 8 + x$). The VCA is better at treating heavily doped system than the rigid band model (RBM) because changes in band structure with carrier doping are calculated self-consistently, whereas the RBM monotonically shifts all bands. We adopt the orbital-orbital interactions obtained by Miyake *et al.* [25]. We take 64×64×4 k-points meshes, 2048 Matsubara frequencies, $T = 0.02$ eV, and an interaction-reducing ratio of $f = 0.415$. For the calculated spin susceptibility [$\chi_s(Q)$] at $x = 0.08$ [Fig. 4(a)], $\chi_s(Q)$ develops at $Q = (\pi, 0)$ corresponding to the wavenumber of stripe-type AFM order $Q^{2D}_{AFM} \sim 1.1$ Å$^{-1}$, in good agreement with the experimental value 1.14 Å$^{-1}$. As $x$ increases, the $\chi_s(Q)$ peak moves towards $(\pi, \pi)$, and the value increases [Fig. 4(b), (c)]. At $x = 0.4$, the peak position is at $Q = (\pi, 0.35\pi) \sim 1.2$ Å$^{-1}$ in Fig. 4 (c), in accord with the experimental value 1.26 Å$^{-1}$. Note that the slight deviations of the calculated values from the experimental $Q$ values are attributable to powder averaging effects, observed and reported elsewhere [26,27].

Figures 4(d)–(i) show calculated Fermi surfaces composed of the Fe-$3d_{YZ, ZX}$ and $3d_{X2-Y2}$ orbitals. The circles around $\Gamma = (0, 0)$ and M = $(\pi, \pi)$ are hole pockets, hereafter called the $\alpha$- and $\gamma$-pockets, respectively, and the ellipse around the X = $(\pi, 0)$ point is an electron pocket (called the $\beta$-pocket). Conventional nesting denotes a momentum transfer at the Fermi surface but "nesting" we use hereafter includes non-zero energy transfer. In all compositions, there are three intra-orbital nestings raising the spin susceptibility: the $\alpha$-$\beta$ nesting within $d_{YZ, ZX}$, and the $\gamma$-$\beta$ and $\beta$-$\beta$ nestings within $d_{X2-Y2}$ denoted by arrows. At $x = 0.08$, both $\alpha$-$\beta$ and $\gamma$-$\beta$ nestings enhance $\chi_s(\pi, 0)$ because the $\alpha$- and $\gamma$-pockets are very similar in size to the $\beta$-pocket. In

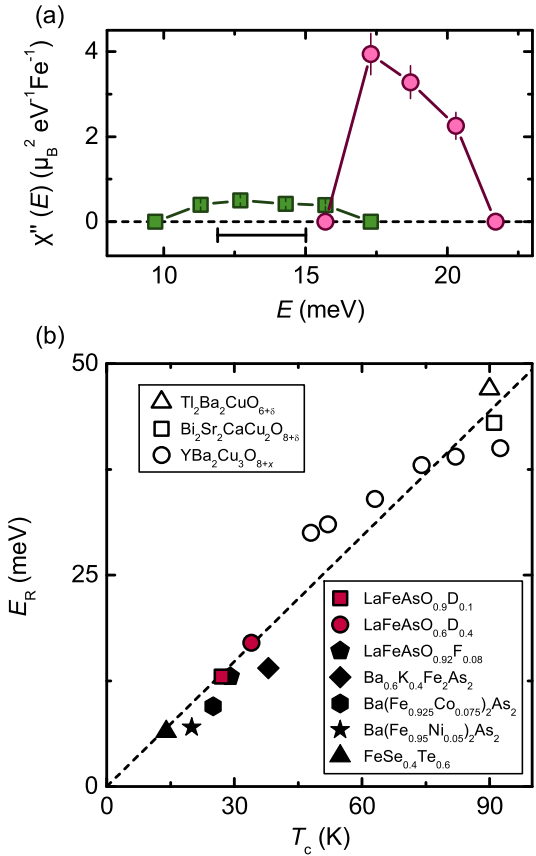

FIG. 3 (color) (a) Energy dependence of $Q$-integrated dynamical spin susceptibility, $\chi''(E)$ at $T = 7$ K and $x = 0.1$ (squares) and 0.4 (circles). The horizontal bar is instrumental resolution. (b) The $T_c$ dependence of the $E_R$ in the iron pnictides (filled symbols) and cuprates (open symbols) [16]. The red square and circles are the present data of the samples with $x = 0.1$ and 0.4, respectively. Dashed line is the averaged slope of 5.7 $k_BT_c$.

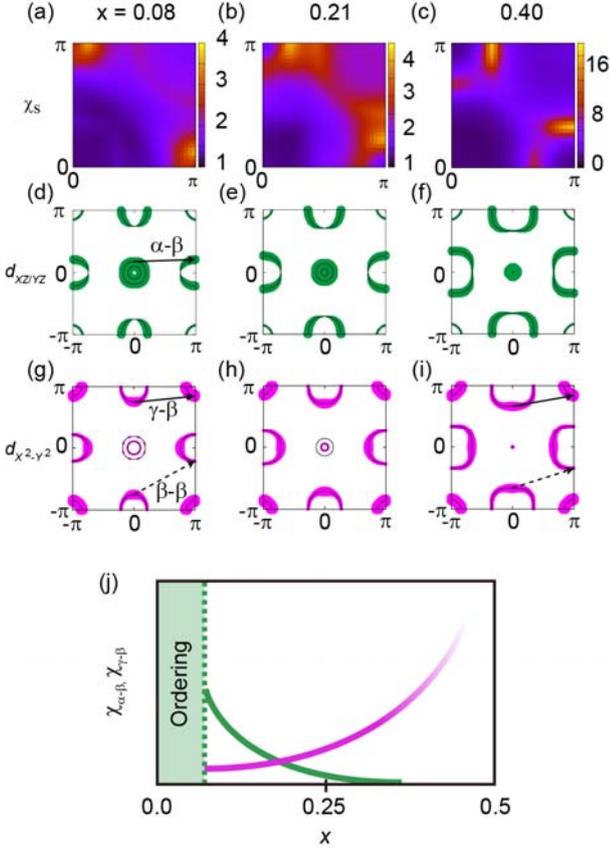

FIG. 4 (color) Contour plot of the spin susceptibility for $x = 0.08$ (a), 0.21 (b), and 0.40 (c). (d)–(i), Fermi surfaces for $x = 0.08$ (d, g), 0.21 (e, h), and 0.40 (g, i). The thickness represents the strength of the $3d_{YZ, ZX}$ (d)–(f) and $3d_{X^2-Y^2}$ (g)–(i) orbital characters of iron. The arrows are nesting vectors. (j) Schematic plots of the $\chi_s$ originating from α-β and γ-β nestings (green and pink lines respectively).

particular, the α-pockets with double pieces strengthen the α-β nesting; the intra-orbital excitations within $d_{YZ, ZX}$ orbital contributes predominantly to the development of the $\chi_s$ peak, where $\chi_{YZ, ZX} / \chi_{X^2-Y^2}$ is 1.3 at $Q = (\pi, 0)$. As $x$ increases, the α-β nesting weakens from the continual shrinking of the α-pocket [see Fig.4(d)–(f)]. These results are close to those obtained by the RBM. In contrast, as $x$ increases, the γ-β nesting becomes stronger because the γ-pocket is unchanged irrespective of $x$ and the β-pocket becomes flat [see Fig.4(g)–(i)]. This flattening also enhances the β-β nesting, and then both γ-β and β-β nestings increase the $\chi_s$ peak at $Q = (\pi, 0.35\pi)$ and $x = 0.40$. To confirm whether the spin fluctuations induce superconductivity, we calculated the eigenvalue of the Eliashberg equation. A previous theoretical study [12] suggested that α-β and γ-β nestings tend to favour the $s\pm$ wave states but β-β nesting can result in $d$ wave states. At least for compositions $0.08 \leq x \leq 0.40$, the eigenvalues for $s\pm$ wave states ($\lambda_s$) have been always higher than that for $d$ wave states ($\lambda_d$); it monotonically increases from $\lambda_s$ of 0.41 at $x = 0.08$ to 1.14 at $x = 0.40$ [28], indicating that the incommensurate γ-β nesting can realize high-$T_c$ superconductivity even at $x = 0.4$.

Figure 4(j) shows schematically the doping dependences for $\chi_s$ originating from α-β ($\chi_{\alpha\text{-}\beta}$) and γ-β nestings ($\chi_{\gamma\text{-}\beta}$). With $0 \leq x \leq 0.05$, the large $\chi_{\alpha\text{-}\beta}$ ($\pi$, 0) induces stripe-type AFM ordering. A slight electron doping breaks the ordering due to the reduction in the α-pocket, and commensurate spin fluctuations at $Q \sim (\pi, 0) \sim 1.1$ Å$^{-1}$ develop for $x > 0.05$. These results are almost the same as those obtained from RBM. Although spin excitations within $3d_{YZ, ZX}$ orbital ($\chi_{\alpha\text{-}\beta}$) weaken with $x$, the present calculations using VCA clarify the enhancement of intra-orbital spin excitations within $3d_{X^2-Y^2}$ orbital ($\chi_{\gamma\text{-}\beta}$). Given the opposite doping dependence of $\chi_{\alpha\text{-}\beta}$ and $\chi_{\gamma\text{-}\beta}$, the suppression of the spin fluctuations and superconductivity at $x = 0.2$ can be understood as switching from $\chi_{\alpha\text{-}\beta}$ to $\chi_{\gamma\text{-}\beta}$. As for the over-doped region ($x > 0.5$), the strong $\chi_{\gamma\text{-}\beta}$ may give rise to spin-ordering states that suppress superconductivity. Here, we would like to point out that occurrence of the two-dome structure is not special in 1111-type iron pnictides. Two-dome structure has recently been reported in the other types, (Tl, Rb, K)$_{1-x}$La$_x$Fe$_{2-y}$Se$_2$ [29] and Ca$_{1-x}$La$_x$Fe$_2$(As$_{1-y}$P$_y$)$_2$ [30]. In both cases, higher-$T_c$ was obtained in the second dome which is far from the parent AFM order as the case of the LaFeAsO$_{1-x}$H$_x$.

Finally, we mention the difference of the dome shape of 1111-type. Kuroki *et al.* have pointed out that increasing the pnictogen height from the iron plane following *Ln*-substitution expands the γ-pocket, which strengthens the γ-β nesting[12]. Therefore, strong γ-β nesting via *Ln*-substitution is expected to fill up the valley between the α-β and γ-β nesting at $x \sim 0.2$ for La-1111. As a consequence, we believe that the $T_c$ valley disappears in *Ln*-1111 (*Ln* $\neq$ La) systems, and the cooperation between α-β and γ-β nestings would produce high-$T_c$.

In summary, we investigated the dynamic spin states under the two superconducting domes in LaFeAsO$_{1-x}$D$_x$ by neutron inelastic scattering techniques, and, in addition to the conventional spin fluctuations with $Q^{2D}_{AFM}$ in $0.1 \leq x \leq 0.2$, observed the development of spin fluctuations with different wavenumber from the parent AFM order under the second dome at $x = 0.4$. The RPA calculations with the VCA to treat precisely the doping effects indicate that the spin fluctuations at $x = 0.1$ are due to intra-orbital nesting within Fe-3$d_{YZ, ZX}$, while the spin fluctuations at $x = 0.4$ originate from intra-orbital nesting within Fe-3$d_{X^2-Y^2}$. These experimental findings and calculations suggest that the orbital multiplicity plays an important role in the doping and / or material dependence of the $T_c$ of the iron pnictides.

This research was supported by the Japan Society for the Promotion of Science through the FIRST program, initiated by the Council for Science and Technology Policy. The inelastic neutron scattering measurements were performed at the BL01 of J-PARC under Project No. 2012A0001(U).

[1] Y. Kamihara et al., J. Am. Chem. Soc. 130, 3296 (2008).
[2] R. Zhi-An et al., Chin. Phys. Lett. 25, 2215 (2008).
[3] C. de la Cruz et al., Nature 453, 899 (2008).
[4] T. Nomura et al., Supercond. Sci. Technol. 21, 125028 (2008).
[5] I. I. Mazin et al., Phys. Rev. Lett. 101, 057003 (2008).
[6] K. Kuroki et al., Phys. Rev. Lett. 101, 087004 (2008).
[7] J.-H. Chu et al., Phys. Rev. B 79, 014506 (2009).
[8] A. F. Wang et al., Phys. Rev. B 85, 224521 (2012).
[9] S. Iimura et al., Nat. Commun. 3, 943 (2012).
[10] T. Hanna et al., Phys. Rev. B 84, 024521 (2011).
[11] S. Matsuishi et al., Phys. Rev. B 85, 014514 (2012).
[12] K. Kuroki et al., Phys. Rev. B 79, 224511 (2009).
[13] T. Moriya et al., J. Phys. Soc. Jpn. 59, 2905 (1990).
[14] The details are discussed in the article: M. Hiraishi et al., Nat. Phys. (submitted)
[15] N. Qureshi et al., Phys. Rev. B 82 184521 (2010).
[16] S. Li and P. Dai, Front. Phys. 6, 429 (2011).
[17] M. M. Korshunov and I. Eremin, Phys. Rev. B 78, 140509 (2008).
[18] T. A. Maier et al., Phys. Rev. B 79, 134520 (2010).
[19] S. Onari et al., Phys. Rev. B 81 060504 (2010).
[20] K. Matan et al., Phys. Rev. B 82, 054515 (2010).
[21] S. Wakimoto et al., J. Phys. Soc. Jpn. 79, 074715 (2010).
[22] P. Blaha et al., An Augmented Plane Wave and Local Orbitals Program for Calculating Crystal PropertiesProperties, Edited by K. Schwarz (Technical University of Wien, Vienna, 2001).
[23] N. Marzari and D. Vanderbilt, Phys. Rev. B 56 12847 (1997); I. Souza et al., Phys. Rev. B 65 035109 (2001). The Wannier functions are generated by the code developed by A. A. Mostofi et al.,: (http://www.wannier.org/)
[24] J. Kuneš et al., Comput. Phys. Commun. 181, 1888 (2010).
[25] T. Miyake et al., J. Phys. Soc. Jpn. 79, 044705 (2010).
[26] M. Ishikado et al., Phys. Rev. B 84, 144517 (2011).
[27] A. E. Taylor et al., Phys. Rev. B 83, 220514 (2011).
[28] Our preliminary fluctuation exchange (FLEX) calculation that takes into account the self-energy, however, shows that the striking increase of $\chi_s$ in $x > 0.3$ is overestimated in the present RPA calculation, but still, even in the FLEX calculation, we find that the eigenvalue does not decrease rapidly with doping as it should in a calculation that assumes the RBM in constructing the model Hamiltonian.
[29] L. Sun et al., Nature 483, 67 (2012).
[30] K. Kudo et al., Sci. Rep. 3, (2013).